\documentclass[useAMS,usenatbib]{emulateapj}
\usepackage{amsmath}
\usepackage{epsfig}
\usepackage{times}

\newcommand{\ltsima} {$\; \buildrel < \over \sim \;$}
\newcommand{\gtsima} {$\; \buildrel > \over \sim \;$}
\newcommand{\lta} {\lower.5ex\hbox{\ltsima}}
\newcommand{\gta} {\lower.5ex\hbox{\gtsima}}

\begin{document}

\title{GRB110721A: An extreme peak energy and signatures of the photosphere}

\author{
M.~Axelsson\altaffilmark{1,2,3,4}, 
L.~Baldini\altaffilmark{5}, 
G.~Barbiellini\altaffilmark{6,7}, 
M.~G.~Baring\altaffilmark{8}, 
R.~Bellazzini\altaffilmark{5}, 
J.~Bregeon\altaffilmark{5}, 
M.~Brigida\altaffilmark{9,10}, 
P.~Bruel\altaffilmark{11}, 
R.~Buehler\altaffilmark{12}, 
G.~A.~Caliandro\altaffilmark{13}
% ...
}

\altaffiltext{1}{Department of Physics, Royal Institute of Technology (KTH), AlbaNova, SE-106 91 Stockholm, Sweden}
\altaffiltext{2}{Department of Astronomy, Stockholm University, SE-106 91 Stockholm, Sweden}
\altaffiltext{3}{The Oskar Klein Centre for Cosmoparticle Physics, AlbaNova, SE-106 91 Stockholm, Sweden}
\altaffiltext{4}{email: magnusa@astro.su.se}
\altaffiltext{5}{Istituto Nazionale di Fisica Nucleare, Sezione di Pisa, I-56127 Pisa, Italy}
\altaffiltext{6}{Istituto Nazionale di Fisica Nucleare, Sezione di Trieste, I-34127 Trieste, Italy}
\altaffiltext{7}{Dipartimento di Fisica, Universit\`a di Trieste, I-34127 Trieste, Italy}
\altaffiltext{8}{Rice University, Department of Physics and Astronomy, MS-108, P. O. Box 1892, Houston, TX 77251, USA}
\altaffiltext{9}{Dipartimento di Fisica ``M. Merlin" dell'Universit\`a e del Politecnico di Bari, I-70126 Bari, Italy}
\altaffiltext{10}{Istituto Nazionale di Fisica Nucleare, Sezione di Bari, 70126 Bari, Italy}
\altaffiltext{11}{Laboratoire Leprince-Ringuet, \'Ecole polytechnique, CNRS/IN2P3, Palaiseau, France}
\altaffiltext{12}{W. W. Hansen Experimental Physics Laboratory, Kavli Institute for Particle Astrophysics and Cosmology, Department of Physics and SLAC National Accelerator Laboratory, Stanford University, Stanford, CA 94305, USA}
\altaffiltext{13}{Institut de Ci\`encies de l'Espai (IEEE-CSIC), Campus UAB, 08193 Barcelona, Spain}
% ...
\label{firstpage}

\begin{abstract}

GRB110721A was observed by the \textit{Fermi Gamma-ray Space Telescope} using its two instruments the Large Area Telescope (LAT) and the Gamma-ray Burst Monitor (GBM). The burst consisted of one major emission episode which lasted for $\sim 24.5$ seconds (in the GBM) and had  a peak flux of $(5.7 \pm 0.2) \times 10^{-5}$ erg s$^{-1}$ cm$^{-2}$. The time-resolved emission spectrum is best modeled with a combination of a Band function and a blackbody spectrum. The peak energy of the Band component was initially $15 \pm 2$ MeV, which is the highest value ever detected in a GRB. This measurement was made possible by combining GBM/BGO data with LAT Low Energy Events to achieve continuous 10\,--\,100 MeV coverage. The peak energy later decreased as a power law in time with an index of $-1.89  \pm 0.10$. The temperature of the blackbody component also decreased, starting from $\sim 80$ keV, and the decay showed a significant break after  $\sim$ 2 seconds. The spectrum provides strong constraints on the standard synchrotron model, indicating that alternative mechanisms may give rise to the emission at these energies.

\end{abstract}


\section{Introduction}

Although the emission mechanisms active in the prompt phase of gamma-ray bursts (GRBs) are still under debate, there is much evidence that the photosphere of the relativistic outflow plays an important role in the formation of the observed spectrum (e.g. Lazzati \& Begelman 2010, Ryde et al. 2010, Guiriec et al. 2011,  Vurm et el. 2011, Giannios 2011, Pe'er et al. 2012). Indeed, a strong contribution from the photosphere was predicted on physical grounds in early works by Goodman (1986) and Paczy\'nski (1986) but this was not considered a viable model since
the observed spectra are, in general, nonthermal. It was, however, realized that  the photospheric emission should be accompanied by nonthermal emission from optically thin regions (e.g. M\'esz\'aros et al. 2002), and that  the photospheric emission can be enhanced and modified from a Planck function by energy dissipation at moderate optical depths (Rees \& M\'esz\'aros 2005). 

The {\it Fermi Gamma-ray Space Telescope} has made observations that support this view. The very bright GRB090902B (Abdo et al. 2009) was observed to have a narrow and steep spectral component that may best be attributed to the photosphere (Ryde et al. 2010, Zhang et al. 2011). At later times during the burst this spectral component widened into a broader feature. This implies that the spectrum of photospheric emission must be able to have a variety of shapes, not only a Planck function (Ryde et al. 2011). A probable explanation for this is the existence of sub-photospheric energy dissipation. Furthermore, many bursts have shown signs of a subdominant photospheric component. For instance, GRB090820A exhibited two spectral peaks: a peak related to a blackbody spectrum with a temperature of  $\sim 40$ keV, and a peak modeled by a Band function (Band et al. 1993) at $E_{\rm p} \sim 1$ MeV (Burgess et al. 2011).

Only a few spectra having peak energies of a few MeV have previously been reported. Gonzalez et al. (2009) presented GRBs observed over the energy range 0.02\,--\,200 MeV, through a joint analysis of data from BATSE and the EGRET TASC on the {\it CGRO}. The maximal value they found was $5 \pm 2$ MeV in GRB981203. Similarly, the PHEBUS experiment onboard {\it Granat}, with an energy  range of 0.1\,--\,100 MeV,  identified spectral breaks in other bursts at around 2.4 MeV in addition to the low energy break \citep{Barat00}. {\it Fermi} observations of GRB090510 showed a peak energy of $\sim 5$ MeV (Ackermann et al. 2010).

In this paper, we study GRB110721A, which besides being very bright  during the first $\sim$ 8 seconds (a few $ \times 10^{-5}$ erg s$^{-1}$ cm$^{-2}$) has an initial peak energy of a  record breaking 15 MeV -- observations made possible by {\it Fermi}'s exceptional spectroscopic capability. Moreover, the $\gamma$-ray emission is dominated by a FRED (fast rise, exponential decay) pulse, which makes this burst ideal to compare with previous studies of such pulses. Apart from the  Band  component,  we identify, with high significance, an additional component which we model using a Planck function. We interpret this as a photospheric component.

\begin{figure}
\begin{center}
\resizebox{0.43\textwidth}{!}{\includegraphics{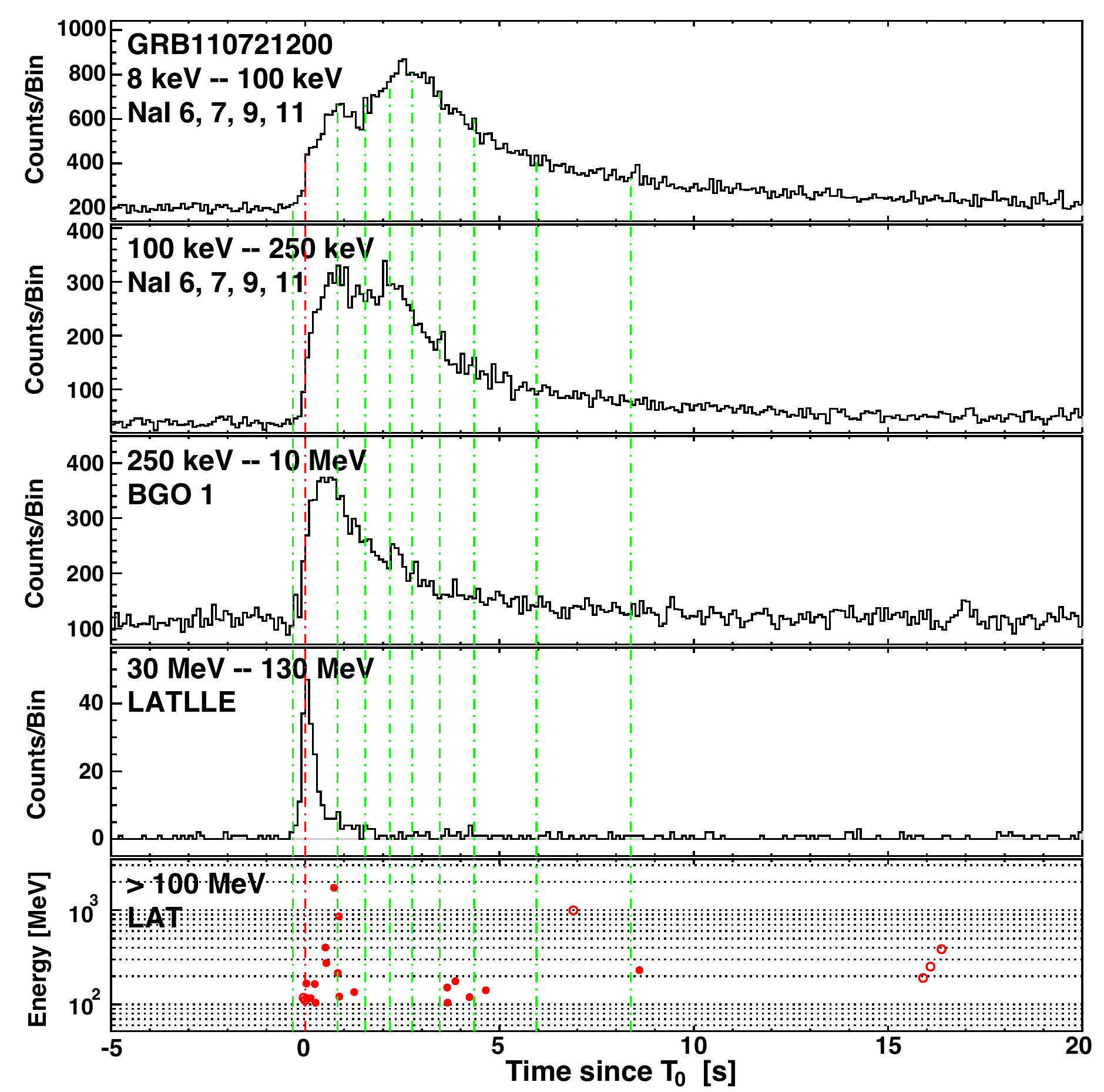}}
\caption{\small{Composite light curve of GRB110721A. (a-d) Count light curves from different energy ranges (NaI  6, 7, 9, 11 and BGO 1 detectors). (e) Individual LAT photons.  
Time intervals are indicated by green lines; the red line shows the trigger time. Filled circles in (e) indicate $>90$\% probability of association with the GRB. The time is relative to the GBM trigger.}}
\label{fig:comp}
\end{center}
\end{figure}

\begin{figure}
\begin{center}
\rotatebox{0}{\resizebox{!}{93mm}{\includegraphics{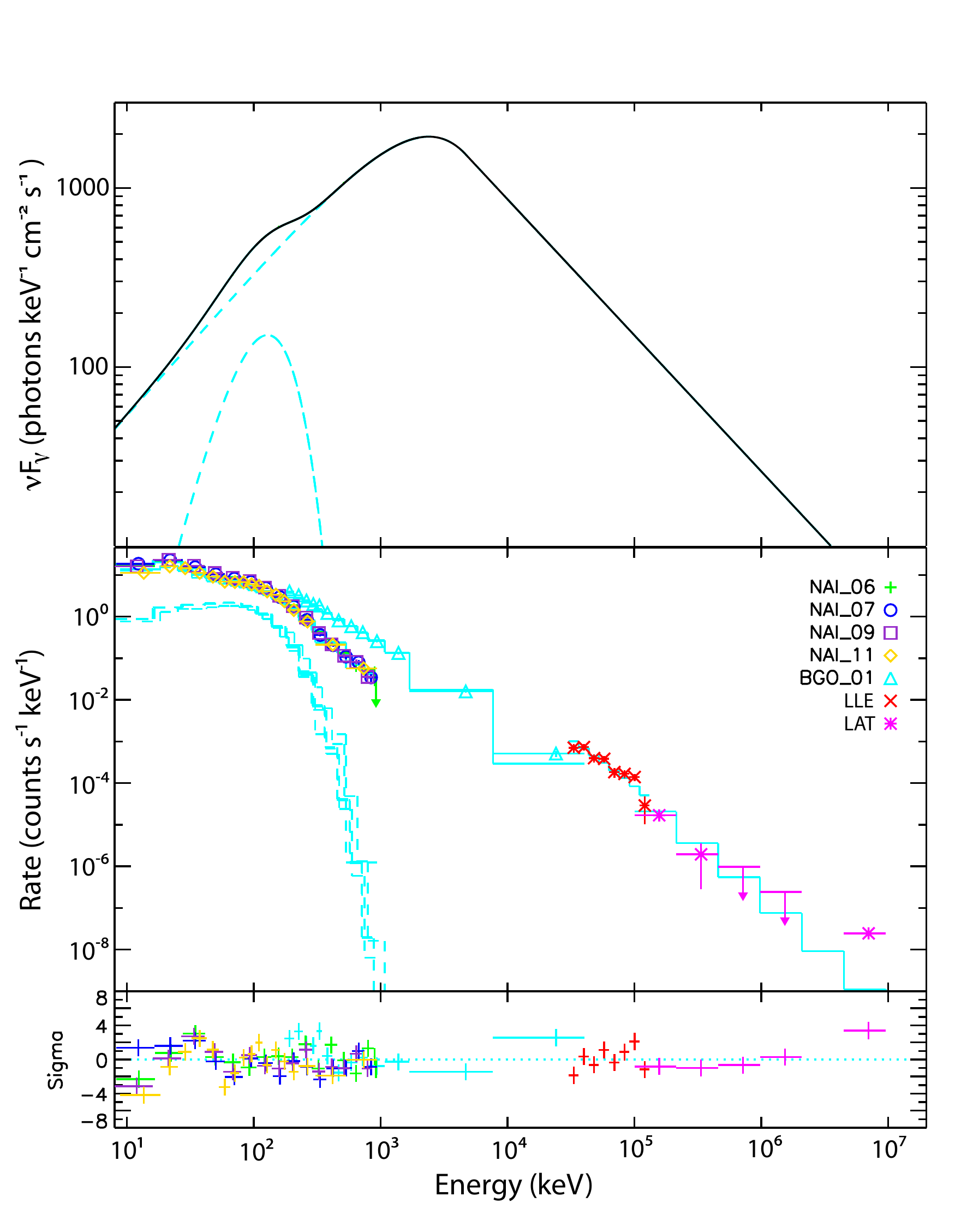}}}
\caption{\small{Spectral fit and residuals  of the time integrated emission ($-0.32$\,--\,8.38 s) and the best fit model, Band + mBB. The top panel shows a $\nu F_\nu$ spectrum, the middle panel a count spectrum and the bottom panel the residuals of the fit.}}
\label{fig:int}
\end{center}
\end{figure}

\setcounter{footnote}{0}

\section{Observations}

On 21 July 2011 the Gamma-Ray Burst Monitor (GBM; Meegan et al. 2009) and the Large Area Telescope (LAT; Atwood et al. 2009) onboard \textit{Fermi} detected high energy emission from GRB110721A (GCNs 12187, 12188). The burst position was triangulated by the Interplanetary Network (IPN), which returned an error box centered in (RA, Dec) = $332.46^\circ$, $-38.63^\circ$ (J2000) and approximately 1.3 $\times$ 0.4 deg wide  (GCN 12195). The intersection of the IPN box with the LAT error circle gives an area of $\sim 1200$ square arcmin centered on (RA, Dec) = $333.2^\circ$, \mbox{$-38.5^\circ$}. We adopted the latter position in our analysis. Figure~\ref{fig:comp} shows a composite count light curve from the various detectors on {\it Fermi}: The NaI detectors (8\,--\,900 keV), the BGO detector (200 keV\,--\,40 MeV) and the LAT (P7V6\_Transient class events, Atwood et al. 2009). The most energetic photon was detected at 4.50\,s, had an energy of $E = 6.3\pm0.6$ GeV and was associated to the GRB with a high ($>0.9$ using the \texttt{gtsrcprob} {\it Fermi} Science tool \footnote{http://fermi.gsfc.nasa.gov/ssc/data/analysis/}) probability. 
The light curve of the burst  is dominated by a single and exceptionally bright FRED-like emission episode with $T_{90} =  24.5$\,s, the time during which 90 \% of the emission is received. The peak of the energy flux  occurs at 0.3 s relative to the GBM trigger with  $(5.7 \pm 0.2) \times 10^{-5}$ erg s$^{-1}$ cm$^{-2}$  (over the energy range 8 keV\,--\,1 GeV). 

In Fig. \ref{fig:comp} we also present the light curve of the LAT Low Energy events (LLE, 30\,--\,130 MeV). The LLE data are produced from a non-standard LAT analysis which for bright sources provides large effective area at low energies, joining the LAT and GBM energy ranges (Pelassa 2011). An observation-specific response matrix is generated for each time range of data using a Monte Carlo simulation of the LAT, which increases the spectral capabilities of the {\it Fermi} data. 
We note that  the LLE light curve for this burst is peculiar since it peaks before, and its duration is significantly shorter than, the GBM light curve, in contrast to what is typically observed in other bursts ({\it Fermi} LAT Collab. in prep.). 
Finally, in Fig. \ref{fig:comp} we explicitly show the NaI light curve in the narrow energy range of  8\,--\,100 keV. Interestingly, at these energies  the light curve differs in that a second peak  appears at $\sim 2$ s. These facts will be discussed further in Sect.~\ref{discussion}.

We have performed a standard analysis with the RMfit 4.0 package (Mallozzi et al. 2005) using TTE data from the NaI  6, 7, 9, 11 and the BGO 1 detectors. We also include the LLE and LAT data.

\section{Spectral Behavior}

A Band function poorly fits the  integrated emission during the pulse ($-0.32$\,--\,8.38 s). The Castor C-statistic (C-stat; Dorman et al. 2003) for the fit is 1078 for 618 degrees of freedom (dof). Based on the earlier detection of a blackbody component in other {\it Fermi} bursts (Ryde et al. 2010, Guiriec et al. 2011, Burgess et al. 2011), we test whether the fit can be improved by including a blackbody. The fit improves significantly to C-stat/dof = 901/616 (corresponding to $> 5 \sigma$ significance). 
Compared to the Band-only fit,  the  peak energy of the Band function shifts from $E_{\rm p} = 1120\pm 60$ keV to $ 2400 \pm 170$ keV.  Since the blackbody component is expected to evolve, (see analysis below and Ryde \& Pe'er 2009) its time-integrated spectrum is expected to be better characterized by a multicolor blackbody (mBB; Pe'er \& Ryde 2011).
Using  such a component instead of the blackbody yields a significant improvement of the fit C-stat/dof= 871/615. Our preferred fit for the time-integrated spectrum is thus the Band + mBB model as shown in Fig. \ref{fig:int}.

We also analyze the time-resolved spectra using data up to 130 MeV (i.e. the LLE data). In order to study the spectral evolution we need a temporal resolution as high as possible. However, since the signal-to-noise ratio (SNR) then also decreases, there is a trade off in our choice of binning. We 
require a SNR of 40 in the most strongly illuminated GBM detector (NaI 9) in the energy range 8.0\,--\,100 keV. This ensures that the spectral fits are well constrained over the energy range where  the blackbody component was found in the time integrated spectrum. This procedure divides the bursts into 8 time bins which are indicated in Table 1.

As for the time-integrated spectrum, we find that the addition of a  blackbody to the Band component improves the fit. In the first 7 bins the C-stat value decreases by a significant amount (see Table 1). Using an assumed Band function spectrum as the null hypothesis, and Band+blackbody as the alternative hypothesis, we determine the signiÞcance of the additional blackbody component. The distribution of the test statistic $\Delta S$ was investigated through Monte Carlo simulations.
We find that an improvement of $\Delta$C-stat = 30 when adding a blackbody component to the fit,  i.e., adding two degrees of freedom, corresponds to a $10^{-7}$ probability of not having a real blackbody component in the spectrum. Similarly, $\Delta$C-stat = 20 corresponds to a $10^{-5}$ probability and $\Delta$C-stat = 10 to a $10^{-3}$ (3$\sigma$) probability. In particular, the normalization of the blackbody component is constrained for the first 7 bins. We tested the robustness of this result by using another time bin selection based on Bayesian blocks (Scargle 1998). The results are fully consistent. We therefore conclude that an extra component is significantly detected in the time-resolved spectra up to $\sim 6$ s. 

\begin{figure}
\begin{center}
\rotatebox{0}{\resizebox{!}{60mm}{\includegraphics{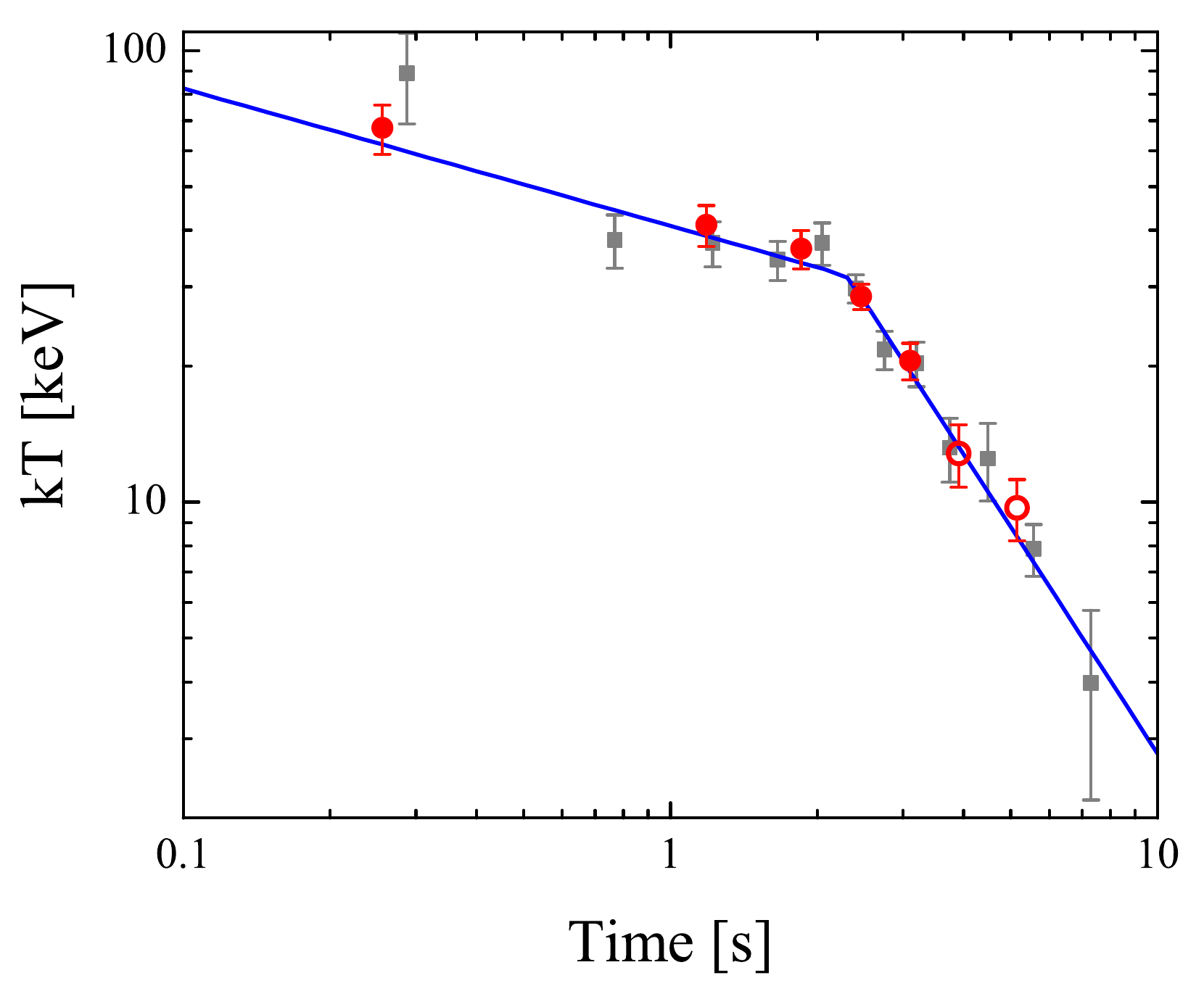}}}
\caption{\small{The blackbody temperature, $kT$, decays as a broken power law (fit function from Ryde 2004). Circles correspond to the binning based on the data below 100 keV, filled/open indicating 5$\sigma$/3$\sigma$ significance of the blackbody component. Gray squares show results from the higher time resolution binning.}}
\label{fig:5}
\end{center}
\end{figure}

In Fig.~\ref{fig:5} we plot the observed temperature of the blackbody as a function of time. First, it is apparent that there is a strong evolution of the temperature. Second, there is a notable break in the decay of the temperature. The energy flux in the blackbody component is approximately 5 \% of the total flux and it peaks at around 2 s, when it reaches 10\%. In the energy band around the peak it is much stronger, and well above the systematic errors (~10\%, Bissaldi et al. 2009). We note that this peak is notably different from the peak of the energy flux pulse, which occurs at 0.3 s, but coincides with the temperature break. Moreover, it coincides with the second peak in the NaI count light curve (Fig. \ref{fig:comp}) which is mainly due to photons below 100 keV. This suggests that this second peak is associated with the blackbody component. The temperature is well fitted with a broken power law (see Ryde 2004) in time with a break at  $2.3 \pm 0.2$ s. Note that such a break does not appear in the decay of $E_{\rm p}$ (Fig.~4). The power law indices before (after) the break are  $-0.30 \pm 0.13$ ($-1.66 \pm 0.15$).  This behavior is quantitatively the same as what was found for the photospheric component in {\it CGRO} BATSE bursts (e.g. Ryde \& Pe'er 2009). 

In order to study the evolution of the Band component we allow for a higher temporal resolution since it is the dominating spectral component. 
We adjust the time bins to provide a minimum SNR of 30 in the counts spectrum from the NaI 9 detector using the full energy range (8.0\,--\,860 keV), which increases the temporal resolution while still maintaining good spectral constraints. Since the burst is initially very strong in the LLE data and spectral evolution is initially very rapid, we divide the first time bin into two. This gives us 13 time bins.

\begin{figure}
\begin{center}
\rotatebox{0}{\resizebox{!}{60mm}{\includegraphics{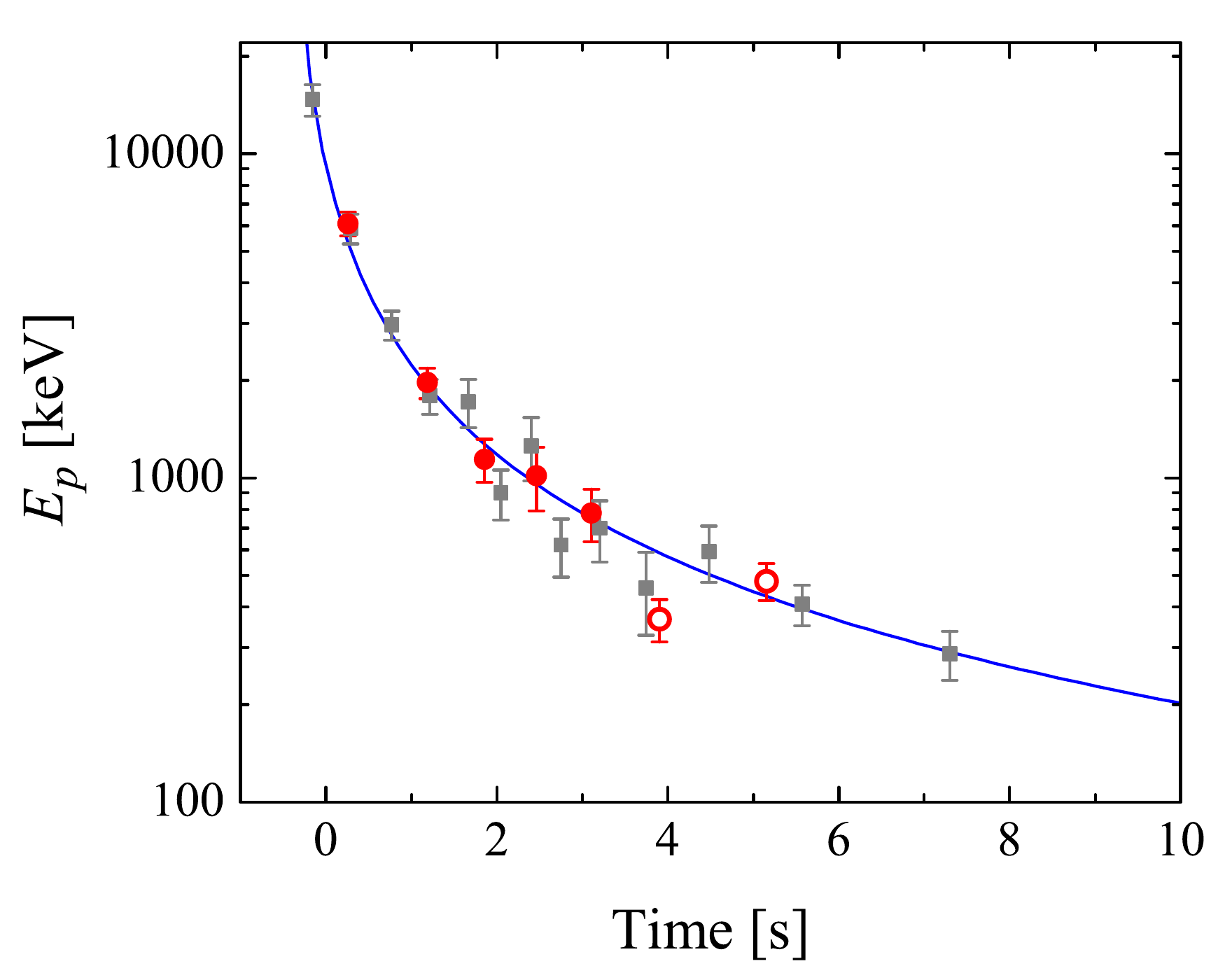}}}
\caption{\small{The evolution of $E_p$ of the Band function component. Symbols as in Fig.~\ref{fig:5}.}}
\label{fig:ep}
\end{center}
\end{figure}

A striking feature in this burst is the unusually high value of the peak energy. The highest value is measured in the first time bin ($-0.32$\,--\,0.0 s). Here the peak energy is $15 \pm 2$ MeV, the low energy slope has a photon index $\alpha = -0.81 \pm 0.08$, the high energy power law has $\beta= -3.5^{+0.4}_{-0.6}$. The fit has C-stat/dof of 679/608. Note that a blackbody component  cannot be constrained by the data in this time bin, but we include it in the following bins. The peak energy of the Band component decreases monotonically with time; 
$E_{\rm p}  = A_{\rm pl} (t-t_0)^d$, where $d= -1.22 \pm 0.13$, and $t_0 = -0.46 \pm 0.10$ s (Fig. \ref{fig:ep}). Attempting to fit the data before and after 2.3\,s (the time where the temperature evolution shows a break) separately gives indices that are compatible within errors. We note that $t_0$ is consistent with the onset of emission in the LLE range .

After a few seconds $E_{\rm p}$ reaches  a few hundred keV, typical for other GRBs (Kaneko et al. 2006). Consistent results are obtained when using other temporal binnings. Figures~\ref{fig:5} and \ref{fig:ep} show the values found for both the coarser (red points) and finer (grey squares) time resolution. The two sets of points are clearly consistent.

The binning used for the time-resolved analysis results in very few data above 100 MeV in each bin. We therefore also studied the time integrated spectra in three broad time intervals in order to increase the signal at the highest energies: $[-0.325, 1.5]$ s,  $[1.5, 5]$ s, $[5, 20]$ s. For each of these bins an extrapolation of the best fit (Band+BB) model of the data below 100 MeV is consistent with the LAT data. The high-energy emission continues for more than 200\, s after the burst trigger ($T_{90}$ for the LAT emission is 253\,s) and decays as a power-law with index $-0.95\pm 0.04$, a typical value for the temporally extended emission observed in other bursts ({\it Fermi} LAT Collab. in prep.).

\begin{table*}
\caption{Fit results using a Band function only and a Band+Blackbody model.}
\begin{center}
\begin{tabular}{llllllllll}
\hline
\hline
{} & \multicolumn{4}{c}{Band function only} & \multicolumn{5}{c}{Band+blackbody} \\
\text{Time (s)} & \colhead{$E_{\rm peak}$ {\rm (keV)}} & \colhead{$\alpha$} & \colhead{$\beta$} & \colhead{\text{C-stat/dof}} & \colhead{$kT$ {\rm (keV)}} & \colhead{$E_{\rm peak}$ {\rm (keV)}} & \colhead{$\alpha$} & \colhead{$\beta$} & \colhead{\text{C-stat/dof}} \\
\tableline
-0.32--0.83 & $ 5410_{-420}^{+410} $ & $ -0.96_{-0.014}^{+0.015} $ & $ -2.82_{-0.08}^{+0.08} $ & $ 660/608 $ & $ 62_{-8.8}^{+10} $ & $ 6490_{-560}^{+500} $ & $ -0.97_{-0.02}^{+0.02} $ & $ -2.9_{-0.09}^{+0.10} $ & 635/606 \\
0.83--1.54 & $ 1330_{-130}^{+120} $ & $ -0.84_{-0.023}^{+0.027} $ & $ -2.71_{-0.06}^{+0.09} $ & $ 741/608 $ & $ 39_{-3.7}^{+4.0} $ & $ 1930_{-220}^{+200} $ & $ -0.87_{-0.03}^{+0.04} $ & $ -2.9_{-0.09}^{+0.11} $ & 696/606 \\
1.54--2.18 & $ 580_{-48}^{+52} $ & $ -0.84_{-0.033}^{+0.035} $ & $ -2.61_{-0.10}^{+0.09} $ & $ 696/608 $ & $ 35_{-3.5}^{+3.4} $ & $ 1140_{-180}^{+220} $ & $ -0.99_{-0.04}^{+0.05} $ & $ -2.9_{-0.17}^{+0.14} $ & 665/606 \\
2.18--2.75 & $ 269_{-22}^{+19} $ & $ -0.68_{-0.05}^{+0.06} $ & $ -2.37_{-0.07}^{+0.06} $ & $ 749/608 $ & $ 30_{-1.9}^{+1.9} $ & $ 1000_{-230}^{+260} $ & $ -1.10_{-0.05}^{+0.06} $ & $ -2.9_{-0.18}^{+0.21} $ & 711/606 \\
2.75--3.46 & $ 344_{-39}^{+55} $ & $ -1.05_{-0.05}^{+0.05} $ & $ -2.46_{-0.16}^{+0.10} $ & $ 657/608 $ & $ 20_{-1.9}^{+1.9} $ & $ 780_{-150}^{+190} $ & $ -1.17_{-0.05}^{+0.06} $ & $ -2.8_{-0.30}^{+0.19} $ & 615/606 \\
3.46--4.35 & $ 309_{-34}^{+41} $ & $ -1.12_{-0.05}^{+0.05} $ & $ -2.32_{-0.07}^{+0.06} $ & $ 682/608  $ & $ 11_{-1.7}^{+2.4} $ & $ 361_{-54}^{+79} $ & $ -1.06_{-0.08}^{+0.09} $ & $ -2.4_{-0.08}^{+0.07} $ & 672/606 \\
4.35--5.95 & $ 456_{-59}^{+64} $ & $ -1.20_{-0.03}^{+0.04} $ & $ -2.64_{-0.08}^{+0.20} $ & $ 633/608 $ & $ 9.3_{-1.1}^{+1.3} $ & $ 453_{-71}^{+62} $ & $ -1.08_{-0.07}^{+0.10} $ & $ -2.7_{-0.07}^{+0.21} $ & 624/606 \\
5.95--8.38 & $ 360_{-46}^{+57} $ & $ -1.13_{-0.05}^{+0.05} $ & $ -2.64_{-0.40}^{+0.16} $ & $ 724/608 $ & $ 4.9_{\textit{unc}}^{+4.3} $ & $ 345_{-61}^{+58} $ & $ -1.10_{-0.12}^{+0.16} $ & $ -2.6_{-0.38}^{+0.17} $ & 724/606 \\
\tableline
\end{tabular}
\end{center}
\end{table*}

\section{Discussion}
\label{discussion}
The value  of $E_{\rm p} = 15\pm 1.7$ MeV found in the initial time bin in GRB110721A is the highest measured in a GRB spectrum. Capturing such a high initial peak energy is a testament to the importance of acquiring good quality BGO and LLE data in bursts; this paper provides the first such realization of this contribution to the GRB paradigm.

Bursts with hard spectra below their $\nu F_{\nu}$ peaks are frequently observed.  Those with indices $\alpha > -1.5$ below this peak cannot possess electrons that radiate synchrotron emission in the expected fast cooling regime, within this spectral window; this is the so-called fast-cooling $\alpha$ index limit (Preece et al. 1998). GRB110721A is just such an example, and its spectrum can be consistent with optically-thin synchrotron emission only if the non-thermal electrons are cooled on timescales much longer than those on which they are injected/replenished.  Demanding that the cooling break should lie well above 15 MeV, combined with the requirement that the observed $\nu F_{\nu}$ peak at 15 MeV in the first time window corresponds to the synchrotron peak, constrains the emission region to be smaller than typical photospheric radius scales (Ryde et al., in prep.), i.e. $10^{10}-10^{13}$\,cm when accounting for GRB bulk motions (e.g., see Rees \&
 M{\'e}sz{\'a}ros 2005).  This is a significant restriction on synchrotron emission models that invoke dissipation zones outside the photosphere. Therefore, it is desirable to entertain alternative explanations for the very ``blue'' emission seen in GRB110721A, e.g. Compton scattering of thermal photons.

The cleanest signature of the emission in GRBs is found in smooth FRED emission pulses \citep{ford95}. Ryde (2004) identified a few such pulses in the {\it Compton Gamma Ray Observatory (CGRO)} BATSE catalogue which were consistent with having a Planck function spectrum throughout their durations over the observed spectral range 25\,--\,2000 keV, suggesting a photospheric origin of the emission. It was further argued in Ryde (2005) that the spectral break in apparently nonthermal spectra can be interpreted as the photospheric emission peak. While the peak is modeled by a Planck function, an additional power law component is needed to account for the nonthermal character of the spectrum. 
More importantly, the temperature of the blackbody in all pulses is observed to have a characteristic, recurring, behavior:  it decreases following a broken power law in time (see further Ryde \& Pe'er 2009). These results suggest that photospheric evolution in GRBs may exhibit well-defined characteristics. 

The evolution of $kT$ seen here is similar to what was observed by  BATSE over individual GRB pulses where a thermal emission component was identified (Ryde 2004).  Moreover, if this burst had been observed by BATSE it would be best modeled by a blackbody and a power law, which is apparent by limiting the energy range to that of BATSE.  GRB110721A therefore confirms the BATSE results on the behavior of the photospheric emission. 

Finally, the two peculiarities of the 30\,--\,100 MeV emission, its early onset and short duration, can be explained by the fact that the emission originates in the same component as detected by the GBM.  This component initially dominates the LLE range due to the exceptionally high value of $E_{\rm p}$ but moves into the GBM range as the burst evolves. 
Therefore the emission detected in the LLE range precedes, and has shorter duration than, the emission seen at lower energies; this mirrors the behavior seen in BATSE bursts with FRED-like light curves (Norris et al. 1996).

In summary, for GRB110721A we have shown that (i) initially the peak energy has an extreme value $E_{\rm p} = 15$~MeV, the identification of which was made possible by
the high quality LLE data. This value combined with the fact that   $\alpha > -1.5$ cannot be explained by the standard optically-thin synchrotron model.  (ii) In addition to the Band function a narrow spectral component  (consistent with a blackbody) is statistically required by the data. (iii) The spectral evolution of this blackbody is similar to what  was found in BATSE pulses having a photospheric component. (iv)  The blackbody component evolves {\it differently} compared to the  Band component. (v) The second peak in the NaI count light curve coincides with the time when the blackbody component flux is the highest. These facts provide strong evidence for the existence of  photospheric emission in GRB110721A.

\section{Acknowledgments}

The {\it Fermi} LAT Collaboration acknowledges support from a number of agencies and institutes for both development and the operation of the LAT as well as scientific data analysis. These include NASA and DOE in the United States, CEA/Irfu and IN2P3/CNRS in France, ASI and INFN in Italy, MEXT, KEK, and JAXA in Japan, and the K.~A.~Wallenberg Foundation, the Swedish Research Council (623-2009-691) and the SNSB in Sweden. Additional support from INAF in Italy and CNES in France for science analysis during the operations phase is also gratefully acknowledged. The {\it Fermi} GBM Collaboration acknowledges support for GBM development, operations and data analysis from NASA in the US and BMWi/DLR in Germany. EM is supported by Carl Tryggers Stiftelse f\"or Vetenskaplig Forskning.


\begin{thebibliography}{}

\bibitem[Abdo et al.(2009)]{2009ApJ...706L.138A} Abdo, A.~A., Ackermann,  M., Ajello, M., et al.\ 2009, \apjl, 706, L138 
\bibitem[Ackermann et al.(2010)]{ack2010}Ackermannm, M. et al. 2010, \apj, 716, 1178
\bibitem[Atwood et al.(2009)]{LAT} Atwood,W. B., Abdo, A. A., Ackermann, M., et al. 2009, ApJ, 697, 1071 
\bibitem[Band et al. (1993)]{band} Band, D., Matteson, J., Ford, L., et al. 1993, ApJL, 413, 281
\bibitem[Barat et al. (2000)]{Barat00} Barat, C., Lestrade, J.~P., Dezalay, J.-P., et al.\ 2000, ApJ, 538, 152 
\bibitem[Bissaldi et al.(2009)]{biss09}  Bissaldi et al. 2009, Experimental Astronomy, 24, 47
\bibitem[Burgess et al.(2011)]{2011ApJ...741...24B} Burgess, J.~M., Preece,  R.~D., Baring, M.~G., et al.\ 2011, \apj, 741, 24 
\bibitem[Dorman et al.(2003)]{Dor03} Dorman, B., Arnaud, K. A., \& Gordon, C. A. 2003, BAAS, 35, 641
\bibitem[Ford et al.\ (1995)]{ford95} Ford, L. A., et~al.\  1995, ApJ, 439, 307
\bibitem[Giannios(2011)]{2011arXiv1111.4258G} Giannios, D.\ 2011, arXiv:1111.4258 
\bibitem[Gonz{\'a}lez et al. (2009)]{gonz09} Gonz{\'a}lez, M., Carrillo-Barrag{\'a}n, M., Dingus, B. et al. 2009, ApJ, 696, 2155
\bibitem[Goodman (1986)]{good86} Goodman, J., 1986, ApJ, 308, L47
\bibitem[Guiriec et al.  (2011)]{2010arXiv1010.4601G} Guiriec, S., et al.\ 2011, ApJ, 727, L33
\bibitem[Kaneko et al. (2006)]{kaneko06} Kaneko, Y., Preece, R.~D., Briggs, M.~S. et al. 2006, 166, 298
\bibitem[Lazzati \& Begelman(2010)]{2010ApJ...725.1137L} Lazzati, D., \& Begelman, M.~C.\ 2010, \apj, 725, 1137 
\bibitem[Malozzi et al.(2005)]{mal05} Mallozzi, R. S., Preece, R. D., \& Briggs, M. S. 2005, RMFIT, A Lightcurve and Spectral Analysis Tool, (Huntsville: Univ. Alabama) 
\bibitem[Meegan et al.  (2009)]{GBM} Meegan, C., Lichti, G., Bhat, P. N., et al. 2009, ApJ, 702, 791
\bibitem[M{\'e}sz{\'a}ros et al.(2000)]{mes02} M{\'e}sz{\'a}ros, P., Ramirez-Ruiz, E., Rees, M. J., \& Zhang, B. 2002, ApJ, 578, 812
\bibitem[Paczy\'nski (1986)]{pac86} Paczy\'nski, B., 1986, ApJL, 308, L43
\bibitem[Pe'er et al.(2011)]{peer11} Pe'er, A., Zhang, B.-B., Ryde, F., et al. 2011, arXiv:1007.2228
\bibitem[Pe'er \& Ryde(2011)]{2011ApJ...732...49P} Pe'er, A., \& Ryde, F.\ 2011, \apj, 732, 49 
\bibitem[Pelassa (2011)]{2011AIPC.1358...41P} Pelassa, V.\ 2011, American Institute of Physics Conference Series, 1358, 41 
\bibitem[Preece et al.(1998)]{1998ApJ...506L..23P} Preece, R.~D., Briggs, M.~S., Mallozzi, R.~S., et al.\ 1998, \apjl, 506, L23 
\bibitem[Rees \& M{\'e}sz{\'a}ros (2005)]{rm05} Rees, M.~J., \& M{\'e}sz{\'a}ros, P. 2005, ApJ, 628, 847 
\bibitem[Ryde (2004)]{ryde04} Ryde, F. 2004, ApJ, 614, 827 
\bibitem[Ryde (2005)]{ryde05} Ryde, F. 2005, ApJL, 625, L95 
\bibitem[Ryde \& Pe'er (2009)]{rydepeer09} Ryde, F. \& Pe'er, A. 2009, ApJ, 702, 1211 
\bibitem[Ryde et al. (2010)]{2010ApJ...709L.172R} Ryde, F., et al.\ 2010,  ApJL, 709, L172 
\bibitem[Ryde et al.(2011)]{2011MNRAS.415.3693R} Ryde, F., Pe'er, A., Nymark, T., et al.\ 2011, \mnras, 415, 3693 
\bibitem[Scargle(1998)]{sca98} Scargle, J.~D. 1998, \apj, 504, 405
\bibitem[Vurm et al.(2011)]{2011ApJ...738...77V} Vurm, I., Beloborodov,  A.~M., \& Poutanen, J.\ 2011, \apj, 738, 77 
\bibitem[Zhang et al.(2011)]{zhang2011} Zhang, B.-B. et al. 2011, \apj, 730, 2
\end{thebibliography}
\end{document}